\documentclass[a4paper,fleqn,usenatbib]{mnras}

\usepackage[T1]{fontenc}
\usepackage{ae,aecompl}

\usepackage{graphicx}	
\usepackage{amsmath}	
\usepackage{amssymb}	

\title[The Galactic bulge]{The origin of stellar populations in the Galactic bulge from chemical abundances}

\author[Matteucci et al.]{F. Matteucci$^{1, 2, 3}$\thanks{E-mail:
matteucci@oats.inaf.it}, V. Grisoni$^{1, 2}$, E. Spitoni$^{4}$, A.
Zulianello$^{1}$, A. Rojas-Arriagada$^{5,7}$, \newauthor M.
Schultheis$^6$, N. Ryde$^{8}$\\
 $^1$ Dipartimento di Fisica, Sezione di Astronomia, Universit\`a di
Trieste, via G.B. Tiepolo 11, I-34131, Trieste, Italy \\
 $^2$ I.N.A.F. Osservatorio
  Astronomico di Trieste, via G.B. Tiepolo 11, I-34131, Trieste,
  Italy\\
 $^3$ I.N.F.N. Sezione di Trieste, via Valerio 2, 34134 Trieste, Italy\\
 $^4$ Stellar Astrophysics Centre, Department of Physics and Astronomy,
Aarhus University, Ny Munkegade 120, DK-8000 Aarhus C,
\\Denmark\\
 $^5$ Instituto de Astrofisica, Facultad de Fisica, Pontificia Universidad
Catolica de Chile, Av. Vicuna Mackenna 4860, Santiago,
\\Chile\\
 $^6$ Universit\`e C$\hat{o}$te d'Azur, Observatoire de la C$\hat{o}$te
d'Azur, CNRS, Laboratoire Lagrange, Bd de l'Observatoire, CS 34229,\\06304
Nice Cedex 4, France \\
$^7$ Millennium Institute of Astrophysics, Av.Vicu$\tilde{n}$a Mackenna 4860, 782-0436 Macul, Santiago, Chile\\
$8$ Lund Observatory, Department of Astronomy and Theoretical Physics, Lund University, Box 43, SE-221 00 Lund
}

\begin{document}
\date{Accepted . ; in original form xxxx}

\pagerange{\pageref{firstpage}--\pageref{lastpage}} \pubyear{xxxx}

\maketitle

\label{firstpage}

\begin{abstract}
In this work, we study the formation and chemical evolution of the Galactic bulge with particular focus on the abundance pattern ([Mg/Fe] vs. [Fe/H]), metallicity
and age distribution functions. We consider detailed chemical evolution models for the Galactic bulge and inner disc, with the aim of shedding light on the connection between these components and the origin of bulge stars. In particular, we first present a model assuming a fast and intense star formation, with the majority of bulge stars forming on a timescale less than 1 Gyr. Then we analyze the possibility of two distinct stellar populations in the bulge, as suggested by Gaia-ESO and APOGEE data. These two populations, one metal poor and the other metal rich, can have had two different origins: i) the metal rich formed after a stop  of $\sim$ 250 Myr in the star formation rate of the bulge, or ii) the metal rich population is made of stars formed in the inner disc and brought into the bulge by the early secular evolution of the bar. We also examine the case of multiple star bursts in the bulge with consequent formation of multiple populations, as suggested by studies of microlensed stars. After comparing model results and observations, we suggest that the most
likely scenario is that there are two main stellar populations, both made mainly by old stars ($> 10$ Gyr), with the metal rich and younger one formed from inner thin disc stars, in agreement with kinematical arguments. However, on the basis of dynamical simulations, we cannot completely exclude that the second population formed after a stop in the star formation during the bulge evolution, so that all the stars formed {\it in situ}.

\end{abstract}


\section{Introduction}

In the last few years, several spectroscopic surveys: Gaia-ESO (Gilmore et al. 2012), APOGEE (Majewski et al. 2017), Argos (Freeman et al. 2012)  and GIBS (Zoccali et al. (2014), as well as photometric (VVVX, which is the extension of the VVV survey, Minniti et al. 2010) surveys  and missions (Gaia mission, Perryman et al. 2001) have been developed in order to study the formation and evolution of the Galactic bulge. The picture for the bulge formation which is arising from these data is rather complex, and still has to be well understood in terms of Galactic chemical evolution models.\\
In particular,  Hill et al. (2011) by observing bulge red clump stars concluded that their distribution is doubled-peaked, with one peak at [Fe/H]=-0.30 dex and the other at [Fe/H]=+0.32 dex, calling the two populations metal poor (MP) and metal rich (MR), confirmed by Uttenthaler et al. (2012).
More recently, Rojas-Arriagada et al. (2017) with Gaia-ESO data and Schultheis et al. (2017) with APOGEE data, concluded that the metallicity distribution function (MDF) in the bulge is indeed bimodal. Zoccali et al. (2017) also confirmed the existence of two main stellar populations with the MP one being more centrally concentrated.
Bensby et al. (2011; 2013; 2017) by studying microlensed dwarfs and subgiant stars found that the bulge metallicity distribution is multi-modal, with at least four peaks corresponding to different star formation episodes occurred 12, 8, 6 and 3 Gyr ago, thus implying the existence of relatively young stars in the bulge. The existence of young bulge stars has been suggested also by Haywood et al. (2016), implying that these stars belong to the inner disc. On the other hand, Clarkson et al. (2011), Valenti et al. (2013), Renzini et al. (2018) and Nogueras-Lara et al. (2018) concluded that most of the bulge stars are quite old ($> 10$ Gyr). In Renzini et al. (2018), from color-magnitude and luminosity functions of the MP and MR populations obtained from HST photometry,  it is concluded that both MP and MR populations are similarly old. Bernard et al. (2018) inferred the history of star formation of the bulge from deep color-magnitude diagrams of four low reddening bulge  regions and concluded that only 10\% of bulge stars are younger than 5 Gyr, but this fraction rises to 20-25\% in the metal rich peak.\\

\begin{table*}
\caption{Input parameters for the chemical evolution models. In the first column, we write the name of the models. In the second column, we indicate whether we consider a continue star formation or a stop in the star formation process. In the third column, there is the star formation efficiency (in Gyr$^{-1}$). Finally, in the last column, there is the IMF. The IMF labelled Kroupa refers to that of Kroupa et al. (1993), the one labelled Calamida refers to the one of Calamida et al. (2015) and finally Salpeter (1955). The label ``$Mg_{Ia}$ normal'' indicates the yield of Mg from SNe Ia by Iwamoto et al. (1999), whereas ``$Mg_{Ia}$ increased'' is the yield artificially increased, as described in the text.}
\label{tab_01}
\begin{center}
\begin{tabular}{c|cccccccccc}
  \hline
\\
 Model &SFR&$\nu$&IMF& $Mg_{Ia}$\\
&&[Gyr$^{-1}$]& & &\\
\\
\hline

A & continue &  25 & Salpeter & $Mg_{Ia}$ normal\\

 \hline

$A^*$ & continue &  25 & Salpeter & $Mg_{Ia}$ increased\\

 \hline

B & stop (50 Myr) & 25 & Salpeter & $Mg_{Ia}$ normal \\

 \hline

C & stop (150 Myr) & 25 & Salpeter & $Mg_{Ia}$ normal\\

 \hline

D & stop (250 Myr) & 25 & Salpeter &$Mg_{Ia}$ normal \\

 \hline

E  & stop (350 Myr) & 25 & Salpeter & $Mg_{Ia}$ normal\\

 \hline

F & multiple stops & 1--3 & Salpeter & $Mg_{Ia}$ normal\\

 \hline

G & continue & 25 & Calamida & $Mg_{Ia}$ normal\\

 \hline

H (disc) & continue & 1 & Kroupa & $Mg_{Ia}$ increased\\

 \hline

\end{tabular}
\end{center}
\end{table*}

From  the theoretical point of view, several scenarios for the bulge formation have been proposed.
Matteucci \& Brocato (1990) first suggested that to reproduce the MDF in the bulge, one should assume a strong and short burst of star formation with the bulk of stars formed in the first 0.5 Gyr, plus an initial mass function (IMF) more top-heavy than the one in the solar neighbourhood, as for example the IMF of
Scalo (1986) derived for local stars. As a consequence of this, they predicted a plateau in the
[$\alpha$/Fe] ratios in bulge stars longer than in the solar vicinity, with a knee close to  [Fe/H]=0.0 dex.
Their prediction was somewhat confirmed by the first data on [$\alpha$/Fe] ratios by Mc William \& Rich (1994).
\\
Wyse and Gilmore (1992) considered various possibilities for the bulge formation, including the model of Matteucci \& Brocato (1990): i) the bulge formed by accretion of extant stellar systems, which by dynamical friction eventually settled in the center of the Galaxy; ii) the bulge formed by accumulation of gas at the center of the Galaxy and evolved independently of the other components of the Galaxy, with either rapid or slow star formation; iii) the bulge formed by accumulation of metal-enriched gas from the thick or thin disc.
\\
Later on, Ballero et al. (2007) presented an updated version of the model by Matteucci and Brocato (1990) and again
concluded that the bulge formed on a very short timescale, of the order of 0.1 Gyr, that the star formation was much
more efficient than in the solar vicinity by a factor of $\sim$20, and that the initial mass function (IMF) was 
flatter than the one adopted for the solar neighborhood.
\\
These conclusions were also supported by the paper of Cescutti and Matteucci (2011), where it was suggested that either a Salpeter or a flatter IMF were required to reproduce the bulge abundance patterns.
\\
Then, Grieco et al. (2012) aimed at explaining the existence of the two main stellar populations observed in the bulge. They concluded that a stellar population forming by means of a classical gravitational gas collapse can be mixed with a younger stellar population created perhaps by the bar evolution.
\\
Several other works have considered that the bulge formed as a result of secular evolution of the inner disc through bar formation and its subsequent bucking into a  pseudo-bulge boxy/peanut (B/P) structure  (Combes et al. 1990; Norman et al. 1996; Athanassoula 2005; Bekki and Tsujimoto 2011; Shen et al. 2010; Debattista et al. 2017; Buck et al. 2017; Fragkoudi et al. 2018), or a mixed scenario where the secular and spheroidal components coexist (Samland and Gerhard 2003; Tsujimoto and Bekki 2012).
\\
The aim of this paper is to study the chemical evolution of the Galactic bulge by means of detailed chemical evolution models in the light of the newest observational data. We will also study the abundance patterns, MDF and age distribution of the Galactic bulge, and compare the observational data with our model predictions in order to constrain the bulge formation and evolution. In particular, we will discuss how the presence of different episodes of star formation, separated by quiescent periods, can produce visible effects on the [$\alpha$/Fe] vs. [Fe/H] relations, and whether we can build a self-consistent scenario which accounts for the MDF shape, the stellar ages and the [$\alpha$/Fe] vs. [Fe/H] relations at the same time.\\
The paper is organized as follows. In Section 2, we present the observational data which have been used to compare with the predictions of our chemical evolution models. In Section 3, we describe the models adopted in this work. In Section 4, we show the results based on the comparison between observational data and model predictions. Finally, in Section 5, we discuss our results and conclusions.

\begin{figure*}
\includegraphics[scale=0.5]{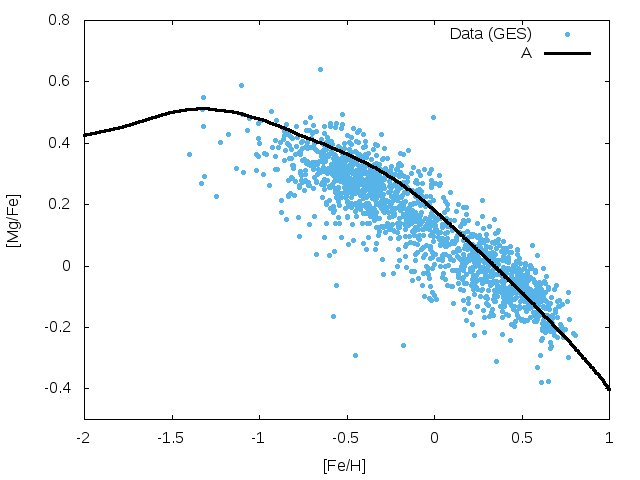}
\includegraphics[scale=0.5]{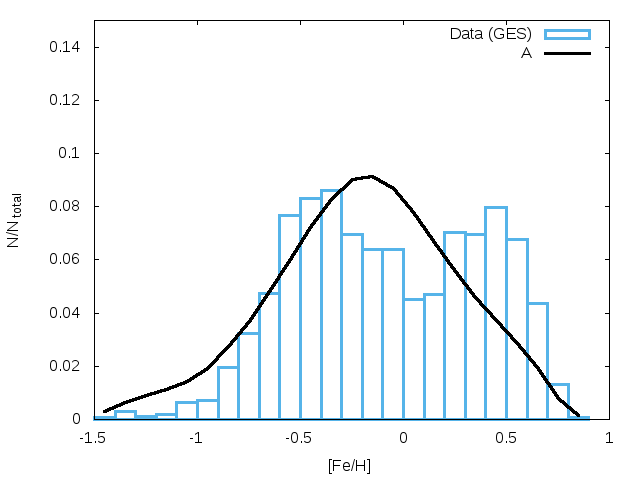}
 \caption{\textit{Left panel}: Predicted [Mg/Fe] vs. [Fe/H] in the Galactic bulge, in the case of Model A (black continuous line), compared with Gaia-ESO data as in Rojas-Arriagada et al. (2017). \textit{Right panel}: Predicted MDF in the Galactic bulge for Model A compared with Gaia-ESO data.}
 \label{fig_01}
\end{figure*}

\section{Observational data}

The observational spectroscopic data that we have used as a comparison to our model predictions are from Gaia-ESO survey (Rojas-Arriagada et al. 2017) and APOGEE (Rojas-Arriagada et al. 2019). In Rojas-Arriagada et al. (2017), 2500 red clump stars were observed in 11 bulge fields: their analysis confirmed the existence of two different stellar populations where the MR one is associated with the boxy/peanut bulge, formed as a result of the secular evolution of the inner disc. We compared our models with both the [Mg/Fe] vs. [Fe/H] and the MDF, found in this paper.
Rojas-Arriagada et al. (2019) analysed the 14th data release from APOGEE data. We have compared again our models with the [Mg/Fe] vs. [Fe/H] relation (DR14) which shows a slightly lower [Mg/Fe] ratio at low metallicity relative to the Gaia-ESO survey data. This can be a problem of different calibrations in the two sets of data. Their MDF is also slightly different from that of Gaia-ESO survey and the existence of the dip indicating two stellar populations is not so evident (see also discussion about differences in the MDF in Schultheis et al. 2017).

Finally, we adopted the ages derived for the bulge stars by Bernard et al. (2018) and Schultheis et al. (2017) by using the CMD-fitting technique, and individual ages based on the CN abundances. Besides finding that 10\% of bulge stars is younger than 5 Gyr, they suggested a fast enrichment rate, in particular $dZ/dt \sim 0.005 Gyr^{-1}$ for the interstellar medium (ISM) in the bulge.
\begin{figure}
\includegraphics[scale=0.35]{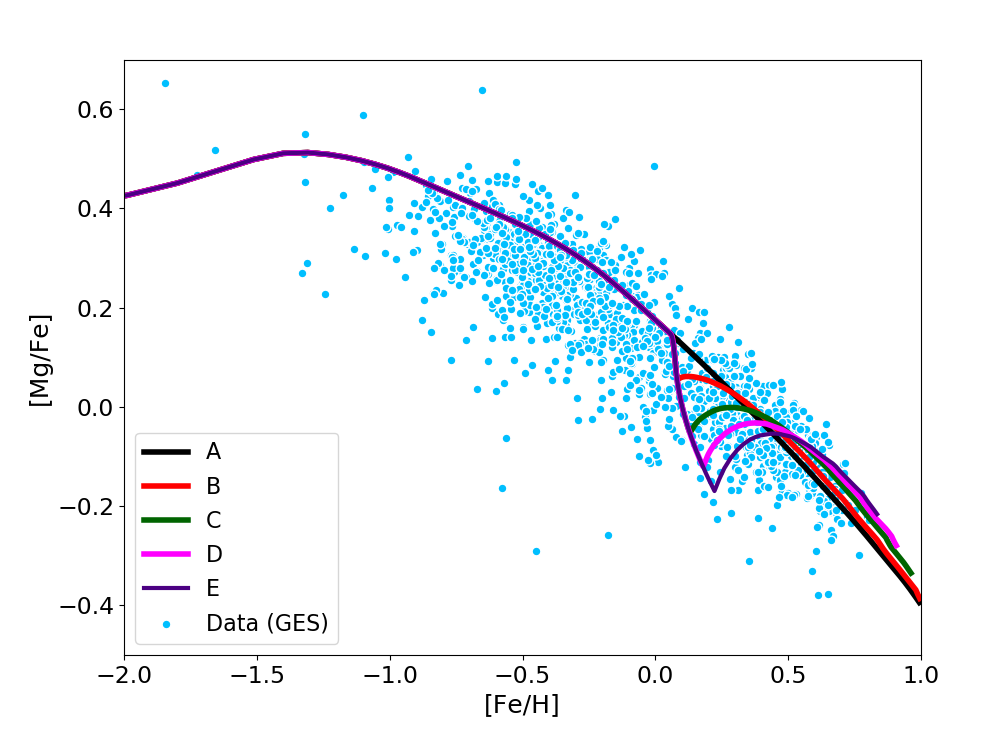}
\includegraphics[scale=0.4]{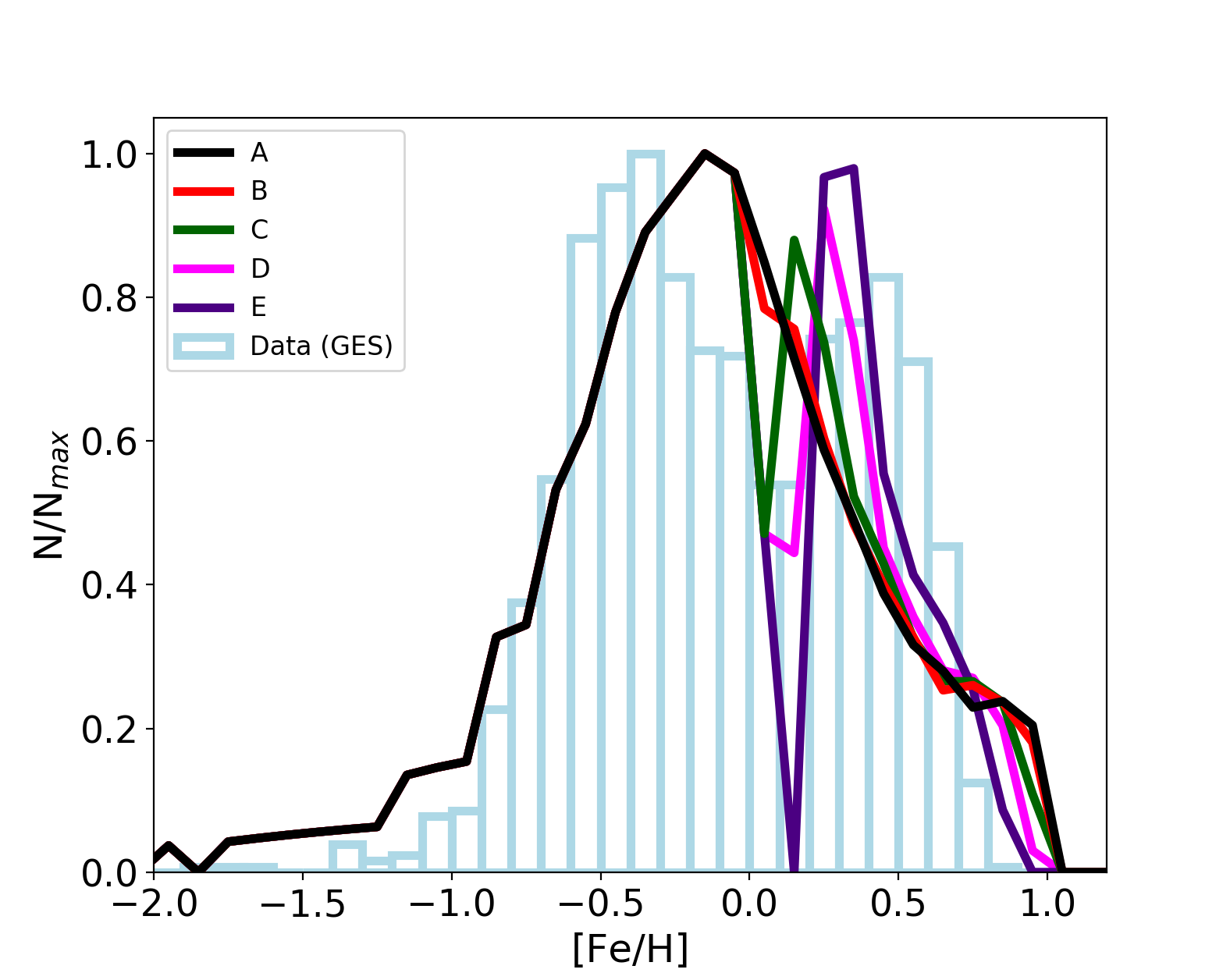}
\includegraphics[scale=0.4]{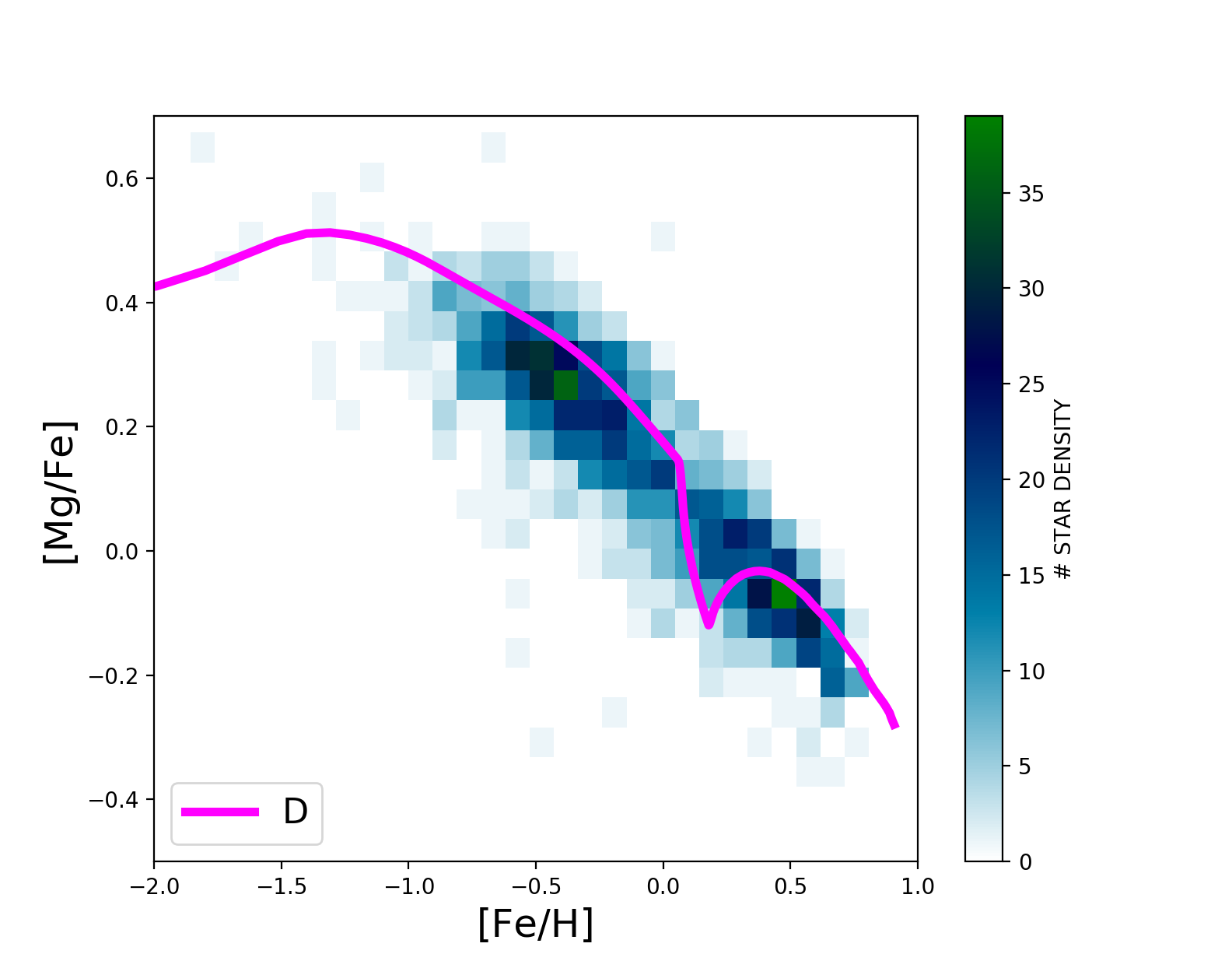}
 \caption{\textit{Upper panel}: Predicted [Mg/Fe] vs. [Fe/H] in the Galactic bulge, in the case of Models A, B, C, D and E with no-stop and stops in the star formation of 50, 150, 250 and 350 Myr, respectively, compared to Gaia-ESO data. \textit{Middle panel}: Predicted MDF in the Galactic bulge for Models A, B, C, D and E compared with Gaia-ESO data. As one can see, longer is the stop in star formation and deeper is the dip between the two populations. The model which best reproduces the data is Model D with a stop of 250 Myr. \textit{Lower panel}: a density plot for the Gaia-ESO data compared to the results of Model D.}
 \label{fig_02}
\end{figure}

\section{The models}

In this work, by means of detailed chemical evolution models we aim at modelling the two stellar populations of the Galactic bulge, the metal-poor (MP) and the metal-rich (MR) ones. We consider two possibilities: i) the MP and MR populations originate from star formation in situ and the MR one forms after a stop in the star formation in the bulge, ii) the MR populations is made of stars originally belonging to the inner disc, whose evolution has been completely disentangled from that of the MP stars.\\
The chemical evolution model for the Galactic bulge that we consider here is the one developed by Grieco et al. (2012, see also Rojas-Arriagada et al. 2017). On the other hand, the chemical evolution model for the Galactic thin disc that we consider here is the one-infall model of Grisoni et al. (2017) (see also Grisoni et al. 2018; Matteucci et al. 2018).\\
We start with a model where the bulge forms by fast gas infall, with a timescale $\tau$= 0.1 Gyr. The assumed gas accretion law is the same for the bulge and disc, but with different timescales of formation:
\begin{equation} \label{eq_01}
\dot G_i(r,t)_{inf}=A(r)(X_i)_{inf}e^{-\frac{t}{\tau}},
\end{equation}
where $G_i(r,t)_{inf}$ is the infalling material in the form of the element $i$ and $(X_i)_{inf}$ the composition of the infalling gas which is assumed to be primordial. The parameter $\tau$ corresponds to the timescale for mass accretion in the Galactic component: as mentioned above, for the Galactic bulge it is assumed to be 0.1 Gyr, whereas for the Galactic thin disc is 7 Gyr in the solar vicinity and it changes with the Galactocentric distance according to the inside-out scenario (Chiappini et al. 2001; Grisoni et al. 2018):
\begin{align} \label{eq_02}
\tau_D \text{[Gyr]}=1.033 \text{r} \text{[kpc]}-1.267,
\end{align}
where $r$ corresponds to the Galactocentric distance; therefore the $\tau$ corresponding to the inner disc (4 kpc) is $\sim 2.7$ Gyr.\\
The quantity $A(r)$ is a parameter fixed by reproducing the present-time total surface mass density in the considered Galactic region.\\
The star formation rate (SFR) is parametrized according to the Schmidt-Kennicutt law (Kennicutt 1998):
\begin{equation} \label{eq_03}
\text{SFR}(r,t)=\nu \sigma_{gas}^k(r,t),
\end{equation}
where $\sigma_{gas}$ is the surface gas density, $k=1.4$ the law index and $\nu$ the star formation efficiency (SFE).\\
The adopted IMF is the Salpeter (1955) one by default for the Galactic bulge and the Kroupa et al. (1993) one for the Galactic disc. We also tested the Calamida et al. (2015) IMF for the Galactic bulge; this IMF was specifically suggested for the bulge stars.\\
Here, we adopt the nucleosynthesis prescriptions of the best model of Romano et al. (2010). However, in one model (Model $A^*$) the yields of Mg from SNe Ia were artificially increased.
This was done for reproducing the observed flattening of [Mg/Fe] at high metallicity, present in the APOGEE data, although this effect is probably artificial (see Nandakumar et al. 2018). In particular, we increased by a factor of 10 the Mg produced in Type Ia SNe. Clearly, this hypothesis is artificial and does not follow what nucleosynthesis models for Type Ia SNe suggest. However, it is interesting to see the effect of increasing the Mg in order to explain the data.
All the models are described in Table 1, where we show the main characterisics of each model: in the first column is the Model name, in column 2 there is the SFR with the indication of  whether the star formation has been halted and for how long, in column 3 is the assumed efficiency of star formation, in column 4 the assumed IMF and finally, in column 5 the assumption about Mg from SNeIa is shown. We show also the inner disc model that we computed  under the hypothesis that the MR population comes from the inner disc, as well as the multiple burst model.

\begin{figure}
\includegraphics[scale=0.4]{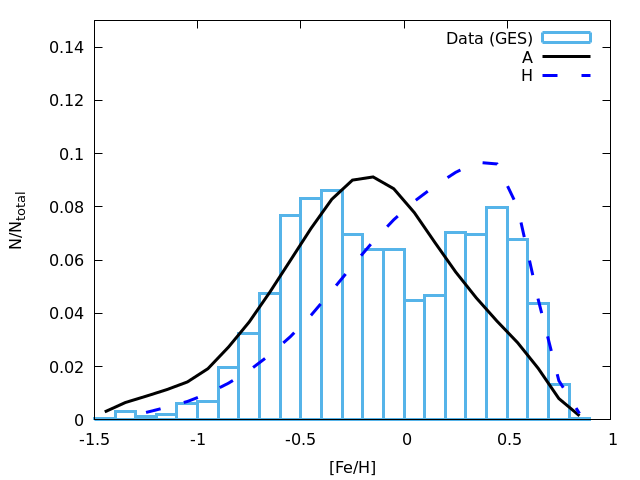}
 \caption{Predicted MDF in the Galactic bulge for Models A (black continuous line) and H ( inner disc population, blue dashed line) compared with Gaia-ESO data. The two peaks, in this case, should be due to the bulge and inner disc populations, respectively.}
 \label{fig_03}
\end{figure}

\section{Results}
Here we show the results for the abundance pattern ([Mg/Fe] vs. [Fe/H]), metallicity distribution function and age distributions as predicted by the various models listed in Table 1.

\begin{figure*}
\includegraphics[scale=0.5]{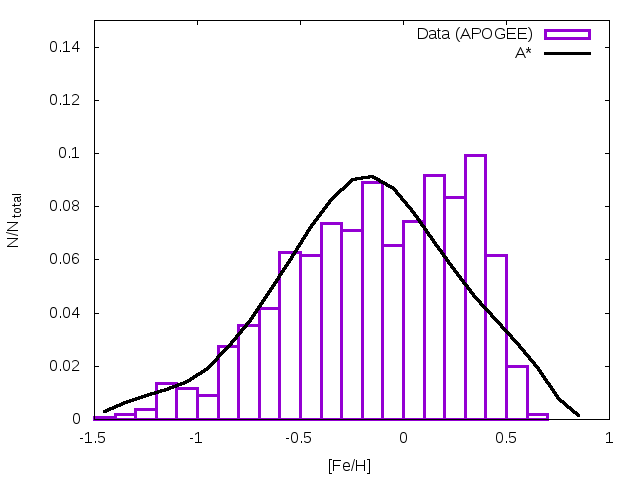}
\includegraphics[scale=0.4]{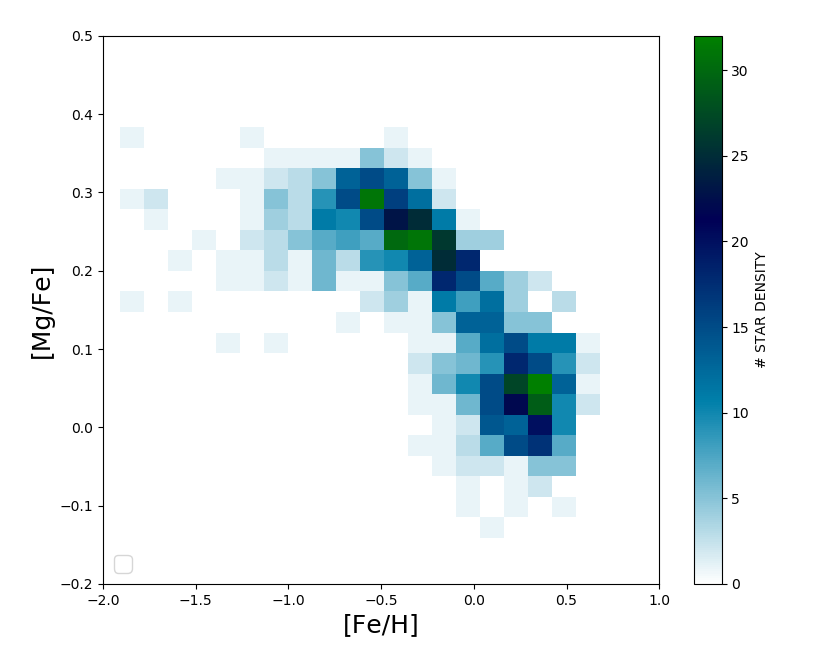}
\caption{\textit{Left  panel}: Predicted MDF in the Galactic bulge for Model $A^*$ (black continuos line) compared
with APOGEE data.\textit{Right panel}: Density plot relative to the APOGEE data for [Mg/Fe] vs. [Fe/H] in the Galactic
bulge.}
 \label{fig_04}
\end{figure*}

\begin{figure*}
\includegraphics[scale=0.35]{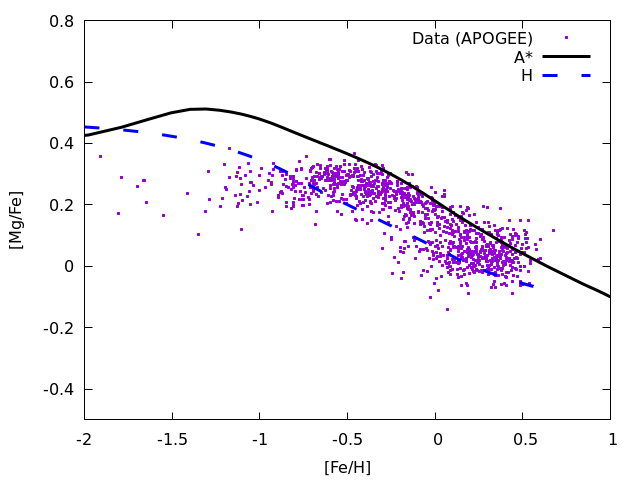}
\includegraphics[scale=0.35]{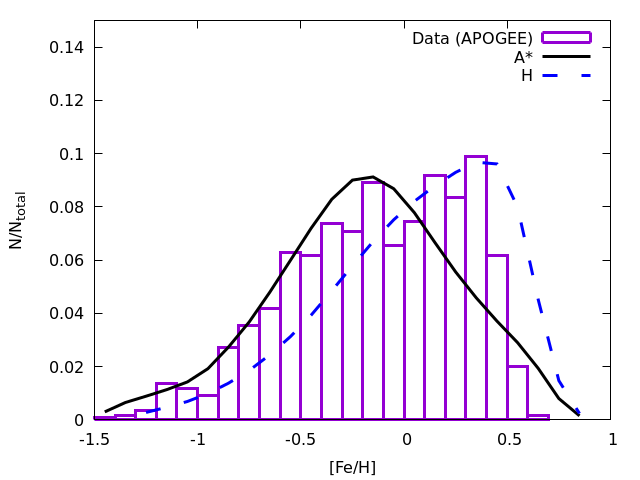}
 \caption{\textit{Left panel}: Predicted [Mg/Fe] vs. [Fe/H] in the Galactic bulge, in the case of Model $A^*$ (black continuous line line) and Model H (blue dashed line) compared to APOGEE data.
 \textit{Right panel}: Predicted MDF in the Galactic bulge for Model $A^*$ and Model H compared with APOGEE data.}
 \label{fig_05}
\end{figure*}

\subsection{Abundance patterns and MDFs}
The first model we started with is the same as in Rojas-Arriagada et al. (2017) and in Grieco et al. (2012) for the MP population: in other words, it is a continuos model characterized by a short and intense star formation burst, typical of classical bulges. This model (Model A in Table 1) can well reproduce the [Mg/Fe] vs. [Fe/H] found by the Gaia-ESO survey, but it does not reproduce well the MDF derived from the same data. In Fig. 1, we show the [Mg/Fe] ratio versus metallicity as well as the MDF predicted by Model A. It is evident that the bimodality observed in the MDF is not reproduced by our Model A, which assumes continuous star formation, therefore we assumed that the star formation stopped during the bulge evolution for a period of time varying from 50 to 350 Myr and we tested the effect that such a halt has on the [Mg/Fe] vs. [Fe/H] relation and the MDF. These models with a stop in the star formation
(Models B, C, D and E) can reproduce the MDF, but produce a hole in the [Mg/Fe] vs. [Fe/H] relation, not immediately visible in the data.
The reason for the occurrence of the hole is that a stop in the star formation determines a stop in the production of Mg, which arises from massive stars, whereas the Fe production continues thanks to SNe Ia which explode, even in absence of star formation, because of their long lifetimes.
To test the existence of such a hole, we have performed a density-plot for the Gaia-ESO data, as shown also in Fig.2. The stellar density plot shows indeed two overdensity regions in correspondance of [Fe/H]=-0.5 and +0.5 dex, respectively,
in agreement with the MDF. Therefore, the hypothesis of a stop in the star formation as the origin of the MR and MP populations cannot be ruled out.
Among the various models, the one which best reproduces the MDF is Model D with a stop of 250 Myr.
However, from the kinematical point of view, the MR population is associated to the Boxy/Peanut X-shaped bulge (Zoccali et al. 2017), while the MP population seems to be distributed isotropically. These facts can support a scenario in which the MR population can originate from the inner disc (e.g. Zoccali et al. 2017)  and not simply from a stop in star formation, although Debattista et al. (2017) have shown that old metal poor stars are dynamically hotter by the time the bar forms and therefore form a weak bar, whereas the more metal rich younger stars are kinematically cooler and therefore form a strong bar with a prominent X-shape, a scenario consistent with a stop in the star formation.  Because of these suggestions, we have then explored also the possibility that the MR population is made of stars of different origin, namely inner disc stars.
\\
To test also this hypothesis,  in Fig. 3 we show the MDF from Gaia-ESO data compared with Models A (for the bulge) and H (for the inner disc). It is worth noting that Model H originates from the thin disc model presented in Grisoni et al. (2018) which reproduces the abundance gradients along the thin disc. The results of Model H can represent the MR population, as shown from the predicted MDF. Model H is devised for the inner thin disc and it assumes an IMF which contains less massive stars than the Salpeter one and is the same as the IMF usually adopted for the solar vicinity (in this case Kroupa et al. 1993). Moreover, Model H assumes a lower star formation efficiency (see Table 1) than assumed for the bulge (see Grisoni et al. 2018).
 We can see from Figure 3 that the disc population can in principle reproduce the second peak in the MDF, and therefore this hypothesis for explaining the MR population appears likely.

\begin{figure*}
\includegraphics[scale=0.35]{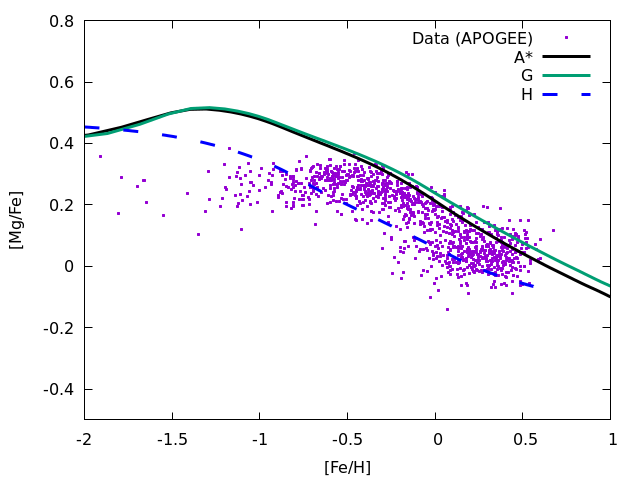}
\includegraphics[scale=0.47]{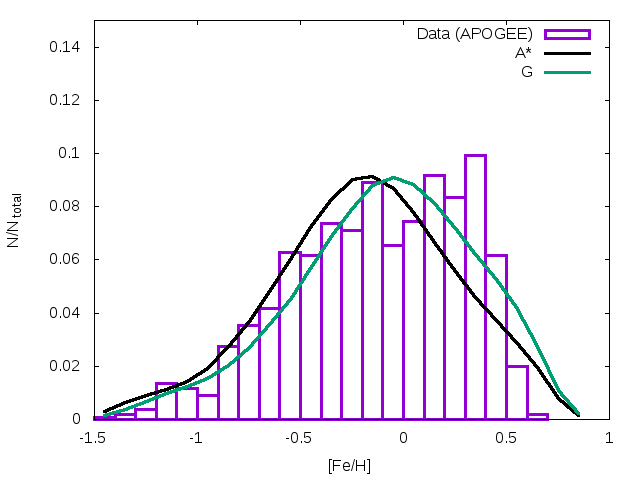}
 \caption{\textit{Left panel}: Predicted [Mg/Fe] vs. [Fe/H] in the Galactic bulge, in the case of Model $A^*$ (black continuous line) and Model G (Calamida et al. 2015 IMF, green line almost overlapping the black  continuous line) and  compared with  APOGEE data.
 \textit{Right panel}: Predicted MDF in the Galactic bulge for Model $A^*$ and G compared with APOGEE data.}
 \label{fig_06}
\end{figure*}

In Fig.4  we show recent APOGEE data (Rojas-Arriagada et al. 2019), and in particular the MDF and the density plot of [Mg/Fe] versus metallicity. The MDF is compared to the results of our Model $A^*$ (Model A with increased Mg yields from SNe Ia).  In Fig. 5 we show the same [Mg/Fe] data as in Fig. 4, compared to the predictions of Model $A^*$.  What we see from these Figures is that  Model $A^*$ seems to overproduce Mg at low metallicities, since these new data have lower [Mg/Fe] ratios at low metallicities; this effect was not present in the comparison with the Gaia-ESO data, as shown in Fig.1, where Model A was fitting very well the observational points. This discrepancy can be due to different calibrations adopted in data reduction for the two different data samples. On the other hand, the increased Mg from Type Ia SNe produces a flatter [Mg/Fe] ratio at high metallicities, in agreement with these data. However, this flattening of the [Mg/Fe] ratio at high metallicity is not yet confirmed and we should be careful in drawing firm conclusions on the yield variation. We are showing this case here only to suggest a possible solution if this trend will be confirmed.\\
As we can see in Fig. 4, these new APOGEE data do not show immediately a clear bimodality in the MDF, as it is instead more evident in Fig. 2, middle panel, for Gaia-ESO data. However, the existence of two separate populations in these data is evident from the [Mg/Fe] density plot in Fig. 4.
\\
Concerning the apparent differences in the MDFs derived from Gaia-ESO and APOGEE data, it should be due to the different spatial regions sampled by the two datasets: in the case of the APOGEE sample, it was selected to contain stars which are close to the plane, with |z|<0.5 kpc. This translates approximately in the stars being located at |b|<4$^{\circ}$ in Galactic latitude. Instead, the Gaia-ESO sample is composed of stars located in a more far-from-the-Galacic-plane region, with -4 > b > -10. As it has been shown by Gaia-ESO and GIBS data, the bulge MDF becomes progressively more dominated by metal-rich stars when we go closer to the Galactic plane. So, in the case of APOGEE data the dip in between the metal-rich and metal-poor peaks is less evident than in Gaia-ESO MDF, because of the larger proportion of metal-rich stars in the APOGEE data sample.
In Fig.5 we show Model $A^*$ and Model H (for the disc) predictions compared to the APOGEE data, both for the
[Mg/Fe] versus metallicity and the MDF. As one can see, in this case the mixture of these two populations provides results in reasonable agreement with both [Mg/Fe] and MDF, so we can conclude that this is an acceptable solution. 
In Fig. 6 we show Model A and G; Model G is identical to Model A except for the IMF which is that of Calamida et al. (2015) instead of the one of Salpeter.
It is evident that the difference between the predictions for the two IMFs is negligible, both in the [Mg/Fe] vs. metallicity
relation and the MDF, and we can conclude that both IMFs are acceptable for describing the bulge stellar populations, with a slight preference for the Salpeter one. Such IMFs require a larger number of massive stars than in the IMFs derived for the solar vicinity, including Kroupa's (2001) IMF.

\subsection{Multiple stops in the star formation}

Bensby et al. (2017), by studying the abundances in microlensed bulge stars, have suggested that
there is a a multi-modal rather than a bimodal MDF in the bulge,  indicating
at least four main stellar populations created in starburst episodes  occurred 12, 8, 6 and 3 Gyr ago.
Although this multi-populations are still to be confirmed, here we have tried to reproduce
this situation by allowing several stops in the star formation rate in our standard Model A, called Model F in Table 1.\\
In particular, in the upper left panel of Fig. 7 we show the predicted SFR as a function of time for Model F;
in this model we have assumed four star formation bursts, with a fixed duration of 250 Myr and separated by long quiescence periods.
A longer burst duration is not likely, because in such a case the bulge stars would form all in the first two episodes.
The star formation efficiency is lower than assumed in Model A. In fact, by assuming $\nu$=25 Gyr$^{-1}$, as in Model A, most of the bulge stars form inside the first 1 Gyr of evolution, so if the star formation occurred in different episodes, distributed over 12 Gyr, the efficiency of star formation during these episodes should have been much lower ($\nu$=1-3 Gyr$^{-1}$). In Model F we assumed a star formation efficiency of 1 Gyr$^{-1}$ in the first burst, whereas in the second, third and fourth burst the efficiency is 3 Gyr$^{-1}$. This choice is rather arbitrary but it allows us to reproduce a situation where an important fraction of young stars is created in the bulge (see next paragraph).
In any case, we have tested that the total mass of bulge stars formed in this model corresponds to that in Model A ($\sim 1.5 \cdot 10^{10} M_{\odot}$).
The presence of multiple star bursts is clearly reflected in the MDF, which appears to show with multiple peaks (see the right upper panel of Fig. 7). The agreement between the observed MDF (Bensby et al. 2017) and the predictions of model F is reasonably good. Finally, we have also checked the effect of the multiple bursts on the abundance pattern, in particular on [Mg/Fe] vs [Fe/H]: what we can see here, is that the predicted track shows holes in correspondance of the stops in the star formation, although they are not so deep as those in Models B, C, D, E. This is due to the lower efficiency of star formation adopted in model F. In fact, a lower efficiency means less stars formed in each burst, so when star formation stops and core-collapse SNe stop exploding, SNe Ia continue to produce Fe thus decreasing the Mg/Fe ratio. The decrease in the [Mg/Fe] ratio
is then lower than in the case where the star formation before the stop has been much higher, with a consequent higher number of SNe Ia produced. In Fig. 7, we show also the density plot for the Bensby et al. (2017) stars. This plot shows that our model predictions are generally following the trend of the data but they are lower than the observations. This is due to the rather low assumed star formation efficiency which produces on average lower [Mg/Fe] ratios, for a given IMF. The lower predicted [Mg/Fe] ratios may suggest that for the bulge a more intense star formation rate should be assumed, in agreement with the previous models, but in such a case most of the stars would form early, thus making the multiburst assumption at variance with the observed abundance ratios.
\begin{figure*}
\includegraphics[scale=0.5]{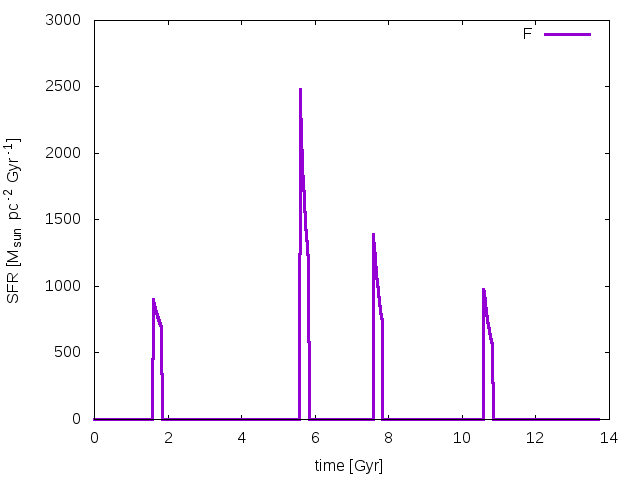}
\includegraphics[scale=0.5]{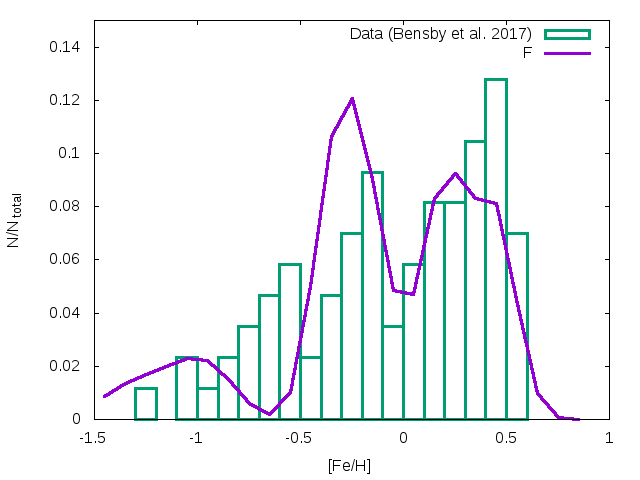}
\includegraphics[scale=0.5]{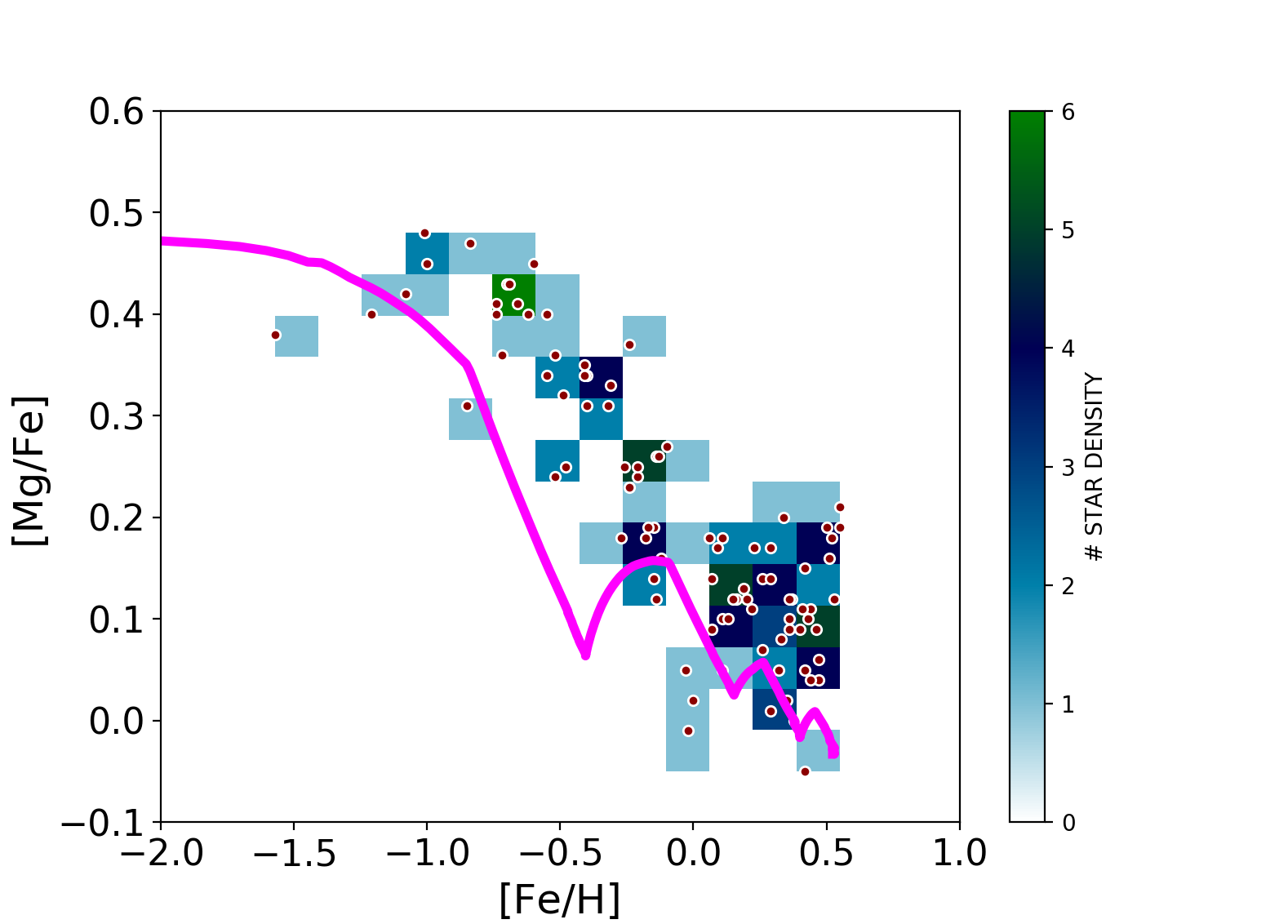}
 \caption{\textit{Upper left panel}: SFR as a function of time, predicted by Model F. \textit{Upper right panel}: Predicted and observed MDF. The predictions are from Model F, the data are from Bensby et al. (2017). \textit{Lower middle panel}: Predicted and observed [Mg/Fe] vs. [Fe/H] in the Galactic bulge, in the case of Model F compared to the data of Bensby et al. (2017). The data are shown as a density plot.}
 \label{fig_07}
\end{figure*}

\subsection{Age distribution}
The ages of the bulge stars can provide a further constraint on the number of stellar populations, although many uncertainties are still present in the derivation of stellar ages. In Figure 8, we show the predicted and observed age distribution in the Galactic bulge. The predictions are
from the various models considered here. The model predictions for the bulge from Models A, D and H are presented as they are computed, as well as corrected by taking into account the errors on the ages obtained by Schultheis et al. (2015; 2017) (with the method of the [C/N] ratio) and by considering only stars not older than 12 Gyr, in accordance with that paper. Schultheis et al. (2015) did not apply any age-cut in their sample. However, due to the limitation of the Martig et al. (2016) method, only 74 stars in the Baade Window do have an age determination. The oldest ages they obtained was 12 Gyr. In our case, in order to consider only stars not older than 12 Gyr we had to remove a large fraction of stars oscillating between $\sim$ 70\%  (Model A) and $\sim$60\% (Model A+H). It is worth noting that we did not shift our model results artificially to lower ages.
To include the observational errors ($\sim$25\%), we followed the approach of Spitoni et al. (2018), as described in their eq. (10). 
At each Galactic time, we added a random error to the ages of the stellar populations formed at Galactic time t. These random errors are uniformly distributed in the interval described by the average errors estimated at that time. 
In Figure 8, we can see that data and model agree remarkably well in the case of Model A, showing that the majority of bulge stars, both from a theoretical and observational point of view, are peaked around an age of 11 Gyr. The peak at 11 Gyr is present also for Model D and Model A+H. The reason why the peak is not at 12 Gyr, as it could be expected, is due to the redistribution and smearing of stellar ages after the cut and the convolution with the observational error; in fact, the bin which includes the age of 12 Gyr spans a range between 11.7 and 12.2 Gyr (see green histograms in Fig. 8), and therefore is affected by the cut at 12 Gyr (stars between 12 and 12.2 have been excluded). This is the reason why this bin contains less stars than bins corresponding to immediately younger ages.
\begin{figure}
 \includegraphics[scale=0.45]{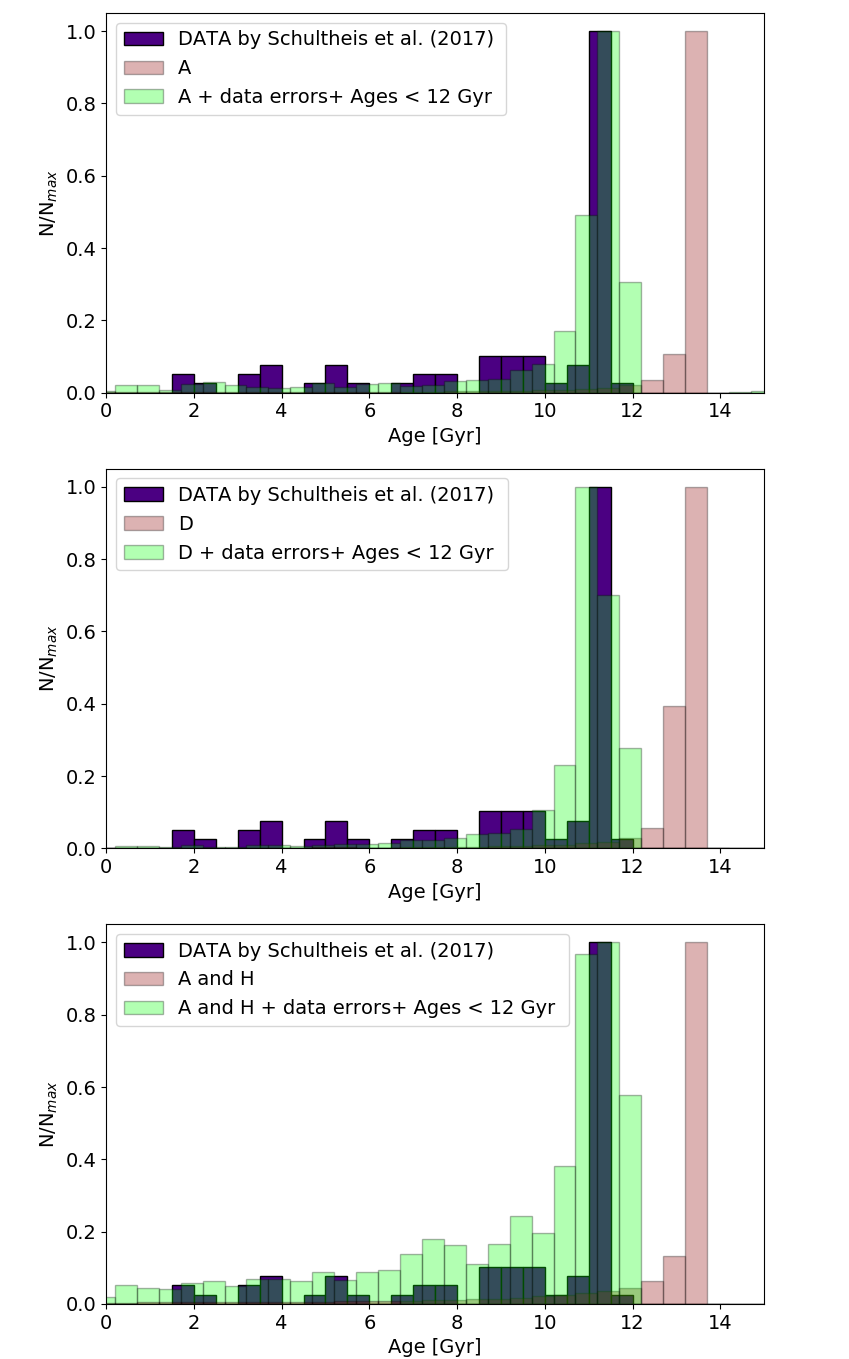}
\caption{Age distribution predicted by the various models, compared to the observational data. {\it Upper panel }: we show the results of Model A with continuous star formation compared to the data of Schultheis et al. (2017)(deep green distribution): the pink distribution represents the  theoretical predictions at a face value, while the light green  distribution is the theoretical one after being convolved with the observational errors; {\it Middle panel}: we show the results of Model D, the colors of the distributions have the same meaning as described for the previous panel; {\it Lower panel}: we show the results of Model A and model H together, convolved with the observational errors. The colors of the distributions have the same meaning as described for the previous panels. }
\label{fig_08}
\end{figure}

In Model A, the predicted number of stars which are younger than 5 Gyr is $\sim 8.7\%$, in agreement with Bernard et al. (2018) who suggest $\sim 10\%$.
In Fig. 8,  we also show the predictions of Model D with a stop in star formation of 250 Myr and therefore with two stellar populations both born in the bulge.  As we can see, the difference relative to  Model A, with only one population, is negligible and the agreement with the data is still quite good, even if more younger stars are produced due to the stop in the star formation. In Figure 8,  we show also the predictions of Model A combined to Model H (for the disc), to test the hypothesis that the MR population can be due to disc stars which formed more slowly than the bulge ones formed in-situ.  In this case, the agreement is also good, since the number of young stars ($<$ 5 Gyr) is $\sim 10\%$, in perfect agreement with Bernard et al. (2018).
\\
Finally, in Figure 9 we show the predictions of Model F with multiple bursts; here, we show the model
predictions after being corrected by the observational errors as quoted by Bensby et al. (2017). 
This model clearly does not show agreement with the Schultheis et al. (2017) data,  so we compared these results with the Bensby et al. (2017) age distribution, from which the suggestion of the multiple bulge populations arose. As one can see in Figure 9, the agreement between our Model F and these data is acceptable when the data are convolved with the errors, and we predict a large fraction of young stars ($<$ 5Gyr) of $\sim 20\%$.
It is worth noting that Bensby et al. (2017) concluded that there are many young stars in the bulge, at variance with other studies, as mentioned in the Introduction.
It is not clear the reason of this discrepancy since the method for deriving ages is similar, namely the isochrone fitting in the CMD. In particular, Bensby et al. (2017) derived the stellar ages by using the Bayesian estimation from isochrones, as described in J{\o}rgensen \& Lindegren(2005). In this method, the
isochrone fitting is done in the luminosity-temperature plane rather than in the CMD.

What arises from these comparisons is that most of the available spectroscopic data on bulge stars suggest that the bulge is formed by a majority of old stars with a minor percentage of truly young stars. Chemical evolution models which well reproduce the [$\alpha$/Fe] ratios in bulge stars need to assume a fast and highly efficient star formation rate which naturally leads to a predominantly old bulge, with the fraction of young stars due either to secular evolution from the inner thin disc or to a stop in the star formation during bulge evolution, since both arguments can be supported by kinematical considerations. From the point of view of age distribution, although many uncertainties are still present, the best model appears the one with the MR population made of inner disc stars, although the other cannot be discarded.

\begin{figure}
\includegraphics[scale=0.30]{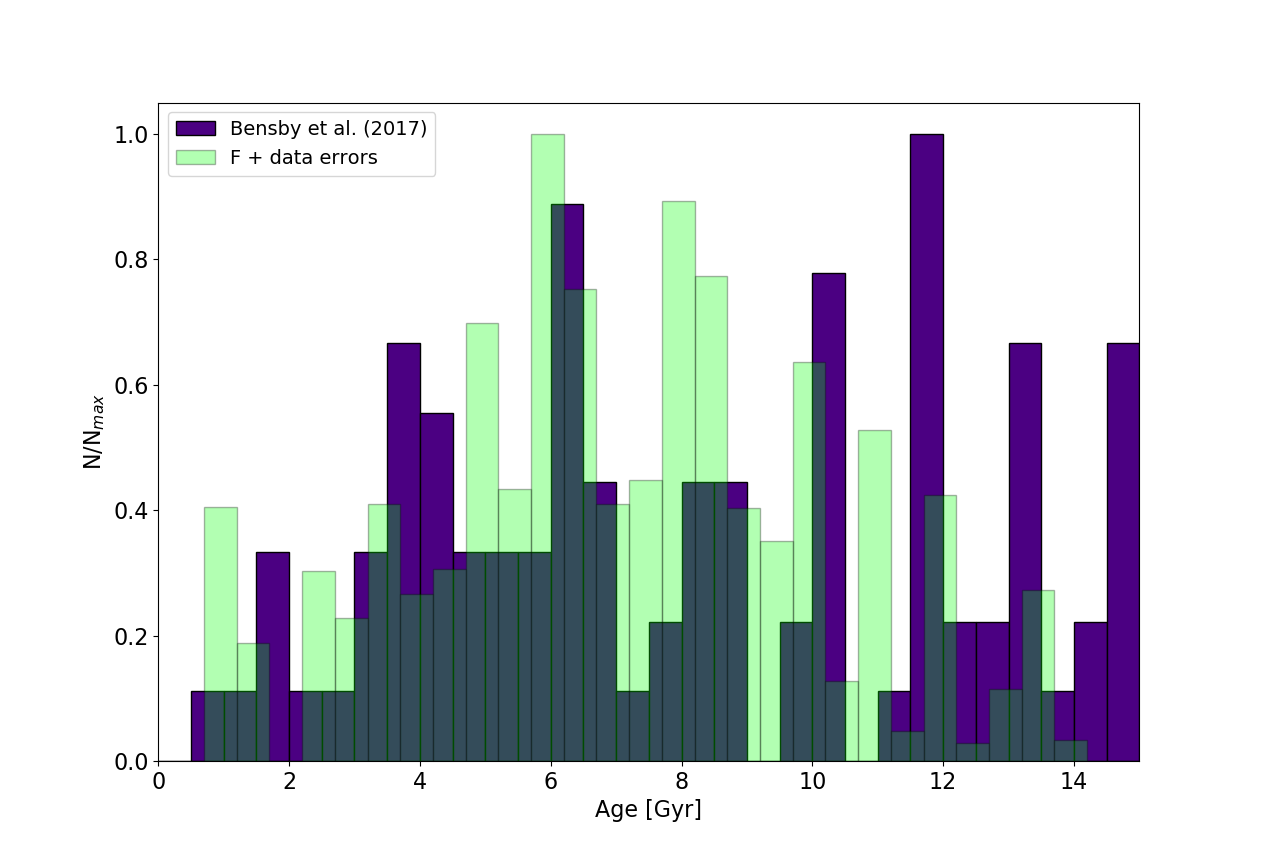}
\caption{Age distribution as predicted by Model F and corrected by the age errors (light green histogram), as described in the text, compared with data of Bensby et al. (2017) (purple histogram).}
\label{fig_09}
\end{figure}

\section{Conclusions}

In this work, we study the formation and chemical evolution of the Galactic bulge with particular focus on the abundance patterns ([Mg/Fe] vs. [Fe/H]), metallicity distribution function and age distribution. We consider detailed chemical evolution models for the Galactic bulge and inner disc, with the aim of shedding light on the formation and evolution of the bulge. In particular, we try to establish if the data can be reproduced by two distinct stellar populations, one metal poor and the other metal rich , and to assess their origin. We explore two main possibilities: i) the two populations have been born in the bulge separated by a period of a stop in the star formation, ii) the MP population was born in the bulge while the MR was formed in the inner disc. We also explore the case of multiple populations born in separate star formation episodes, as suggested by Bensby et al. (2017).
In all the studied cases, except this last one, the MP population forms very quickly  (less than 500 Myr) and with high star formation efficiency (25 Gyr$^{-1}$). The same prescriptions are adopted for the MR one if we assume that it is born in the bulge after a halt in the star formation process. On the other hand, in the multiple burst case the efficiency of star formation during different episodes is assumed to be much lower (from 1 to 3 $Gyr^{-1}$) and the bulge formed on a much longer timescale (several Gyrs).
Finally, in the case where MR population is formed by inner disc stars, the efficiency of star formation is low and typical of the thin disc (1$Gyr^{-1}$).

After comparing model predictions and observational data we can draw some conclusions, summarized as follows:

\begin{itemize}
\item Models with two main stellar populations in the bulge best fit the most recent data from Gaia-ESO and APOGEE. In particular, if the two populations have formed as a result of a stop in the star formation of $\sim 250$ Myr, occurred at early times, one can reproduce the MDF, the [Mg/Fe] ratios and the age distribution of bulge stars. However, this scenario could be  inconsistent with stellar kinematics suggesting that the MR stars are belonging to the B/P X-shaped structure of the bulge, whereas the MP stars are distributed isotropically (Zoccali et al. 2017), although other studies (Debattista et al. 2017; Buck et al. 2017) do not exclude the possibility of explaining the X-shape only with stars formed {\it in situ}.

\item A metal rich population originating from the inner thin disc seems a good suggestion, in the light of the available data. Also in this case, in fact, we can reproduce the MDF, the [Mg/Fe] ratios and the age distribution.

\item The flattening of the [Mg/Fe] ratio at high metallicity in the last APOGEE data could be reproduced by assuming a larger Mg production from SNe Ia. However, this flattening is not present in all the existing  bulge data and therefore we cannot draw firm conclusions on this point.

\item The assumed Salpeter IMF can well reproduce the data and the results differ negligibly from those obtained with Calamida et al. (2015) IMF derived for the bulge. Therefore, we confirm that the bulge IMF should be flatter in the domain of massive stars than the Scalo (1986), Kroupa et al. (1993) and Kroupa (2001) IMFs derived for the solar vicinity. 

\item The results of a multiple burst regime with the bursts occurring from 3 to 12 Gyr ago, as suggested by Bensby et al. (2017), can roughly reproduce their data but  is in conflict with all the other data and predict a large fraction of young bulge stars which is not found in the majority of the other studies. In addition, a multiple burst scenario is also inconsistent with the kinematical information.

\item Therefore, we conclude that the bulge overall is old and that both the MP and MR populations contain very old stars. The young stars (10 \% with ages $< $5 Gyr) belong either to the inner disc stars or they have formed {\it in situ} after a stop in star formation no longer than 250 Myr.  The bulge formed the majority of its stars in the first 0.5 Gyr of its evolution, in agreement with most of the previous studies (Matteucci \& Brocato, 1990; Ballero et al. 2007; Cescutti\& Matteucci 2011; Grieco et al. 2012).

\end{itemize}

\section*{Acknowledgments}

V.G. and F.M. acknowledge financial support from the University of Trieste (FRA2016). E.S. acknowledges support from the Independent Research Fund Denmark (Research grant 7027-00096B).
A.R-A. acknowledges partial support from FONDECYT through grant 3180203.This work has been supported by the Ministry for the Economy, Development,
and Tourism's Programa  Iniciativa  Cient\'\i  fica  Milenio through  grant
IC120009,  awarded  to Millenium Institute of Astrophysics (MAS), the BASAL
CATA Center for Astrophysics and Associated Technologies through grant
PFB-06,  and from FONDECYT Regular 1150345. We finally thank an anonymous referee for careful reading and useful suggestions.

\end{document}